\documentclass[sigconf]{acmart}

\usepackage{booktabs}

\usepackage[ruled]{algorithm2e} 

\SetAlFnt{\small}
\SetAlCapFnt{\small}
\SetAlCapNameFnt{\small}
\SetAlCapHSkip{0pt}
\IncMargin{-\parindent}

\AtBeginDocument{%
  }

\copyrightyear{2025}
\acmYear{2025}
\acmConference[SIGGRAPH Conference Papers ’25]{Special Interest Group on Computer Graphics and Interactive Techniques Conference Conference Papers ’25}{Auguest 10--14, 2025}{Vancouver, Canada}
\acmISBN{979-8-4007-1540-2/25/08}

\setcopyright{rightsretained}

\acmDOI{10.1145/3721238.3730750}

\setcitestyle{square}

\usepackage{algpseudocode}
\usepackage{amsmath}
\usepackage{pdfpages} 

\begin{document}
\title{Large-Scale Multi-Character Interaction Synthesis}

\author{Ziyi Chang}
\email{ziyi.chang@durham.ac.uk}
\orcid{0000-0003-0746-6826}
\affiliation{%
  \institution{Durham University}
  \city{Durham}
  \country{United Kingdom}
}

\author{He Wang}
\email{he_wang@ucl.ac.uk}
\orcid{0000-0002-2281-5679}
\affiliation{%
  \institution{University College London}
  \city{London}
  \country{United Kingdom}
}

\author{George Alex Koulieris}
\email{georgios.a.koulieris@durham.ac.uk}
\orcid{0000-0003-1610-6240}
\affiliation{%
  \institution{Durham University}
  \city{Durham}
  \country{United Kingdom}
}

\author{Hubert P. H. Shum}
\authornote{Corresponding author.}
\email{hubert.shum@durham.ac.uk}
\orcid{0000-0001-5651-6039}
\affiliation{%
  \institution{Durham University}
  \city{Durham}
  \country{United Kingdom}
}


\renewcommand{\shortauthors}{Chang et. al.}

\begin{abstract}
Generating large-scale multi-character interactions is a challenging and important task in character animation. Multi-character interactions involve not only natural interactive motions but also characters coordinated with each other for transition. For example, a dance scenario involves characters dancing with partners and also characters coordinated to new partners based on spatial and temporal observations. We term such transitions as coordinated interactions and decompose them into interaction synthesis and transition planning. Previous methods of single-character animation do not consider interactions that are critical for multiple characters. Deep-learning-based interaction synthesis usually focuses on two characters and does not consider transition planning. Optimization-based interaction synthesis relies on manually designing objective functions that may not generalize well. While crowd simulation involves more characters, their interactions are sparse and passive. We identify two challenges to multi-character interaction synthesis, including the lack of data and the planning of transitions among close and dense interactions. Existing datasets either do not have multiple characters or do not have close and dense interactions. The planning of transitions for multi-character close and dense interactions needs both spatial and temporal considerations. We propose a conditional generative pipeline comprising a coordinatable multi-character interaction space for interaction synthesis and a transition planning network for coordinations. Our experiments demonstrate the effectiveness of our proposed pipeline for multi-character interaction synthesis and the applications facilitated by our method show the scalability and transferability.
\end{abstract}

%
%



\begin{teaserfigure}
\centering
    \includegraphics[width=\linewidth]{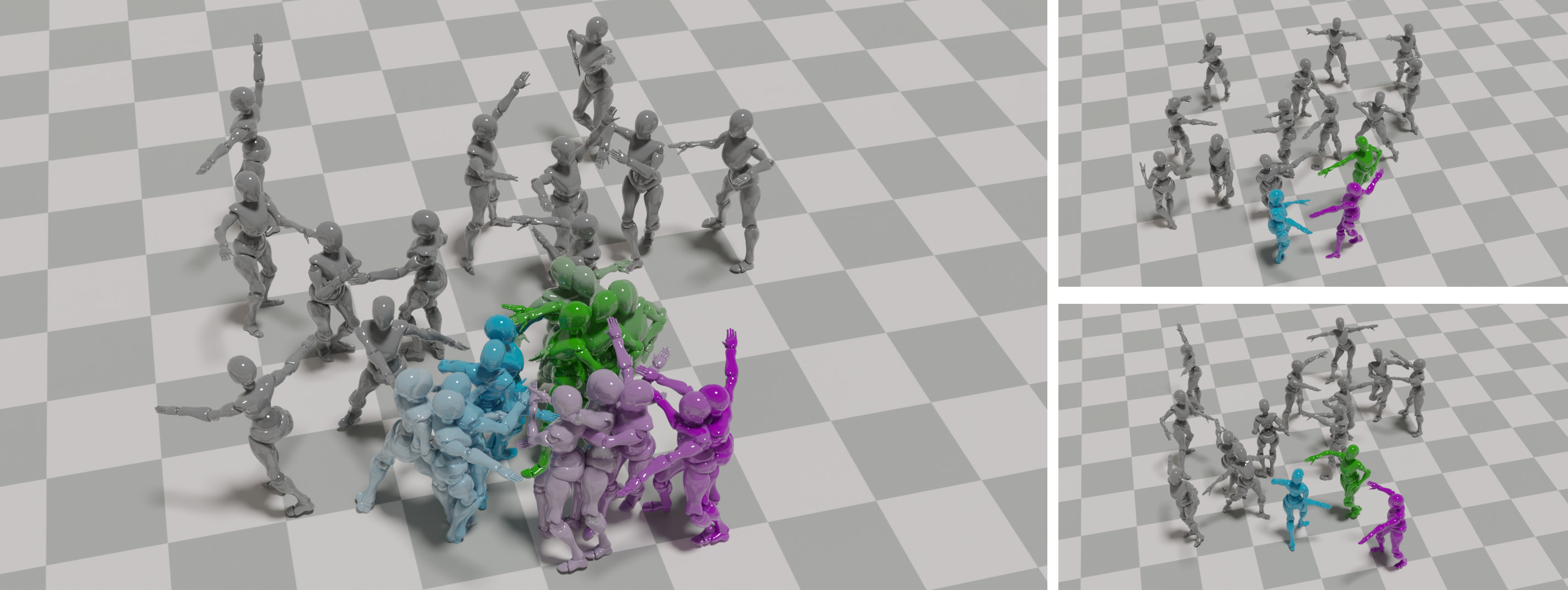}
    \caption{Multi-character interactions coordinated with transition planning. (Left) We highlight the three currently interacting characters with blue, purple, and green, while others are grey. The more saturated the color, the more recent the frame. (Upper right) The key frame of the transition where the blue and purple characters proceed to have a coordinated interaction. (Lower right) The key frame of the transition where the blue and green characters proceed to have a coordinated interaction.}
    \Description{Multi-character interactions coordinated with transition planning. (Left) We highlight the three currently interacting characters with blue, purple, and green, while others are grey. The more saturated the color, the more recent the frame. (Upper right) The key frame of the transition where the blue and purple characters proceed to have a coordinated interaction. (Lower right) The key frame of the transition where the blue and green characters proceed to have a coordinated interaction.}
    \label{fig:teaser}
\end{teaserfigure}

\maketitle

\section{Introduction}

Generating large-scale multi-character interactions is a challenging and important task in character animation. For example, a dance scenario involves characters dancing with partners and simultaneously planning potential interactions with new partners based on spatial and temporal observations. We term such behaviors as \textit{coordinated interactions} in multi-character settings, and decompose them into interaction synthesis and transition planning. The former requires a multi-character interaction space to generate realistic interactions, and the latter demands the ability to plan transitions among characters. As the number of characters can vary, such a method needs to be scalable to support broader applicability.

Multi-character interaction synthesis is an underdeveloped area despite recent advances in character animation, interaction generation, and crowd simulation. Single-character animation focuses on modeling motions for a single character with control \cite{starke2022deepphase}. However, it does not learn the interaction patterns that are critical for multiple characters. Interaction generation with deep learning focuses more on learning motions for two characters \cite{liang2024intergen}. While some optimization-based interaction synthesis methods \cite{shum2008interaction} can produce interactions for multiple characters, they depend on manually designing objective functions that may not generalize well. Crowd simulation involves more than two characters, but their interactions are relatively sparse \cite{charalambous2023greil} and passive \cite{yue2024human}. Our task of large-scale multi-character interaction synthesis requires active and denser interactions with transition planning, and needs to be scalable to multiple characters.

Large-scale multi-character interaction synthesis faces two challenges. The first challenge is the lack of data. Existing datasets for interactions \cite{liang2024intergen} focus on two characters and do not consider coordinated interactions. Existing datasets for crowd simulation \cite{zhong2022data} do not contain dense and close interactions. Capturing such a dataset for our task would be time-consuming and labor-intensive, which becomes unmanageable as the number of characters scales up. The second challenge is to plan dense and close interactions based on spatial and temporal context for multiple characters. Scheduling suitable interactions for multiple characters is a highly correlated problem. In the temporal domain, previous coordination could heavily influence the interactions that follow, and in the spatial domain, the difficulty of planning increases with the increasing number of characters.

We propose a generative pipeline for large-scale multi-character interaction synthesis, including a coordinatable interaction space to generate natural interactions for multiple characters and a transition planning network to coordinate potential transitions among multiple characters. In the absence of suitable data, we divide multi-character interactions into several two-character groups that are modeled by a pre-trained two-character interaction diffusion model. With such a division, we can generalize the natural two-character interaction manifold to an interaction space for multiple characters in the absence of data. Additionally, our division is agnostic to the number of characters and is scalable to multiple characters. To plan transitions for coordinated interactions, we propose a planning network to predict high-level transition plans that are represented as re-grouping choices. Thus, our transition planning network is motion-agnostic, which is transferable to other types of motion. In the absence of suitable data, we propose to train our method through reinforcement learning. Specifically, our coordinatable interaction space works as the environment to generate coordinated interactions, and the transition planning is the policy network. We define transition smoothness and transition diversity as rewards.

\begin{figure*}[htbp]
  \centering
  \includegraphics[width=\linewidth]{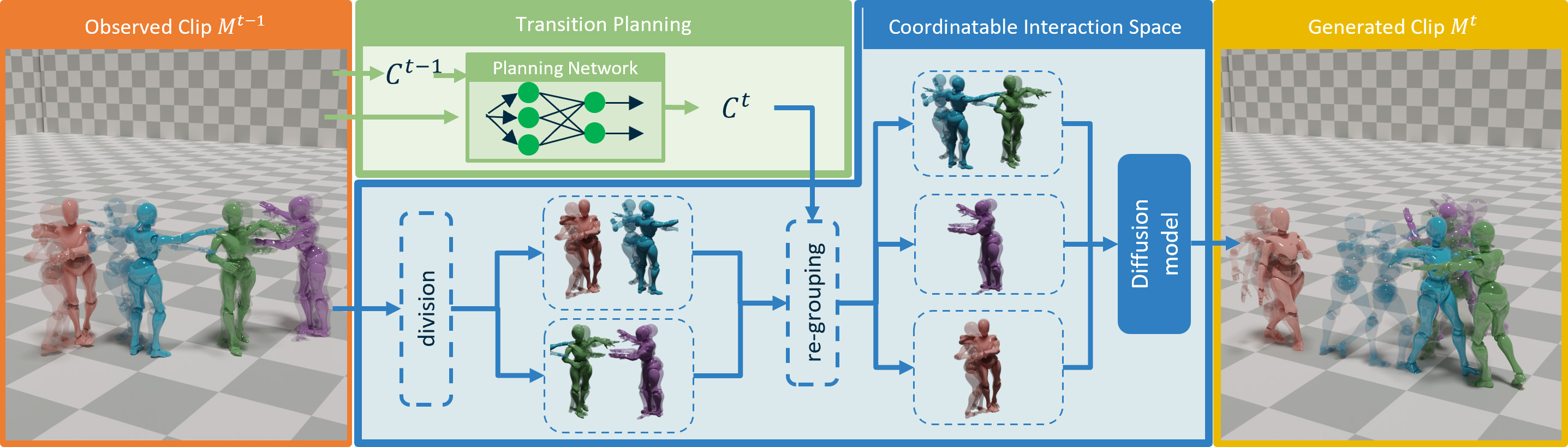}
  \caption{Framework overview. Our pipeline is an autoregressive conditional generative model to plan transitions and synthesize interactions for multiple characters. It has two components: The first component divides multiple characters into groups and leverages a pre-trained diffusion-based model to autoregressively generate interactions for each group. The second component predicts a transition plan based on the observed interactions and serves as the conditional signal for the interaction synthesis.}
  \Description{Framework overview. Our pipeline is an autoregressive conditional generative model to plan transitions and synthesize interactions for multiple characters. It has two components: The first component divides multiple characters into groups and leverages a pre-trained diffusion-based model to autoregressively generate interactions for each group. The second component predicts a transition plan based on the observed interactions and serves as the conditional signal for the interaction synthesis.}
  \label{fig:overview}
\end{figure*}

We train our method with a two-character dancing subset from the InterHuman dataset \cite{liang2024intergen} and evaluate it by synthesizing a larger number of characters and transferring to other motion types. We calculate transition smoothness and hip distance as two metrics for comparison. Our experiments show that our method is scalable and transferable.

We summarize our two contributions below:
\begin{itemize}
    \item We propose a framework to synthesize large-scale multiple characters by decomposing their coordinated interactions into interaction synthesis and transition planning.
    \item We propose a method of combining a pre-trained two-character diffusion model and a transition planning network to learn the coordinated interactions via deep reinforcement learning without the requirement for data.
\end{itemize}

\section{Related Work}

\subsection{Deep Learning-based Motion Synthesis}

Deep learning has been extensively utilized for single-character motion synthesis, emphasizing high-fidelity motion details \cite{zhou2023multi} with motion control. Action labels have been popular \cite{xu2023actformer,chang23unifying} to generate various categories of actions. To achieve finer controllability, texts have been explored as a control signal \cite{petrovich2022temos}. Text-to-motion synthesis aims to integrate language representation into pose representation. For example, a unified text-motion joint space has been modeled \cite{tevet2022motionclip} for better representation. More recently, many methods leverage diffusion models \cite{tevet2022human} and design conditional mechanisms to control the generation. They either fine-tune a diffusion model on conditional data \cite{xie2024omnicontrol} or leverage classifier guidance for post-hoc control without requiring conditional data. Other types of control signals, such as physical constraints \cite{yuan2023physdiff} and objects,\cite{li2023object} have also been explored. While the aforementioned methods are effective for modeling character motions, they primarily focus on motion synthesis for a single character. Adding control to such methods does not generalize well to multi-character interactions.

\subsection{Interaction Synthesis}

A popular line of work is based on optimization. It optimizes individual character motions with spatial-temporal constraints \cite{kwon2008two,liu2006composition} and game theory \cite{shum2007simulating,shum2008simulating,shum2010simulating} to connect motions smoothly. Optimization is usually based on motion patches, where a patch includes a short interaction of multiple characters \cite{shum2008interaction,kim2012tiling,won2014generating,yersin2009crowd}. However, a goal function usually needs to be designed manually to perform optimization and may not be well generalized to different motion patterns.

Another line of interaction synthesis is based on modeling the interaction distribution with deep learning methods. The interaction is further defined differently. Some methods simplify interaction synthesis as reaction synthesis \cite{xu2024regennet} where the interaction includes active characters performing actions and passive characters responding to such actions. Some methods synthesize character interactions with a large interaction dataset that is usually designed to have two characters by default. They learn the individual motion distribution by modeling the dependency between two characters \cite{liang2024intergen,xu2023actformer}. Another line of work synthesizes future interactions based on the observed interactions \cite{xu2023joint,guo2022multi}. They extract interaction relationships from observed motions and predict future interactions. Others leverage deep reinforcement learning to learn movements for two characters \cite{zhang2023simulation,won2021control}. However, these methods focus more on two characters and cannot be easily extended to multi-character interactions.

\subsection{Crowd Simulation}

Crowd simulation methods can generally be categorized into two streams: macroscopic and microscopic methods \cite{pelechano2016simulating}. Macroscopic methods recognize characters as active matter. They aim to synthesize the density of such active matter in both the spatial and temporal domains. By such an abstraction, they do not focus on the different motions of individual characters.

Microscopic methods for crowd simulation focus on synthesizing individual movements for multiple characters. Trajectories are widely used to represent the movements of crowds. Many recent methods use reinforcement learning to learn instantaneous actions such as the velocity or acceleration of each character. These methods usually focus on the navigation of crowd characters \cite{charalambous2023greil,panayiotou2022ccp} by designing rewards for collision avoidance and target reaching. Some learn spatial-temporal features from large trajectories of crowd characters \cite{xiang2024socialcvae,mao2023leapfrog}. Human skeletal joint trajectories have also been considered \cite{jeong2024multi} in a manually crafted dataset where several independent groups are concatenated in a scene and their motions are rather static. Although these methods consider more than two characters, their interactions among characters are sparse in time and space.

\section{Methodology}

\subsection{Problem Formulation}

Our goal is to generate large-scale multi-character interactions without available data. In the absence of data, we decompose the coordinated interactions into interaction synthesis for multiple characters and transition planning for future interactions. The former requires a multi-character interaction space to synthesize natural interactions, while the latter demands the ability to plan transitions with close and dense interactions for multiple characters.

We represent interactions with full-body poses and motion clips. We learn full-body poses for more realistic synthesis because full-body poses may provide social cues \cite{fiore2013toward} for realistic interaction synthesis and transition planning. Moreover, to consider coordinated interactions in the temporal domain, we represent motions by short clips that contain a few consecutive frames to provide contextual information that could reflect character intentions for transitions when compared with poses in a single frame.

Specifically, we denote multi-character interactions as
\begin{equation}
    M^{1:T}_{1:N}=\left[M_{1:N}^1,M_{1:N}^2,\cdots,M_{1:N}^t,\cdots,M_{1:N}^T\right],
\end{equation} where $T$ is the total number of motion clips and $N$ is the total number of characters. The $t$-th clip $M^t_{1:N}=\left[m_{1}^t,m_2^t,\cdots,m_n^t,\cdots,m_N^t\right]$ consists of motions for $N$ characters where each clip contains $w$ frames and $m_n^t$ is the $t$-th clip of the $n$-th character. We follow the representations in \cite{liang2024intergen} and represent a motion clip by global positions, local rotations, and velocities of each joint.

\subsection{Pipeline Overview}

Our pipeline is an autoregressive conditional generative model to synthesize coordinated interactions for multiple characters. We design our method to be autoregressive to allow characters to dynamically plan their future transitions based on the observed interactions between themselves and other characters. This is more flexible than planning everything ahead.

Overall, our framework $\mathcal{F}$, as illustrated in Fig. \ref{fig:overview}, consists of two components for synthesizing multi-character coordinated interaction. Specifically, the first component, namely the coordinatable interaction space, provides a natural interaction space for multiple characters to synthesize interactions with coordination signals. The second component, namely transition planning, predicts transition plans to coordinate interactions that arise from the first component.

The first component, as shown in Fig. \ref{fig:approximation}, divides multiple characters into several groups of two characters. In the absence of data, such a division allows us to simplify the multi-character interaction space by means of learnable two-character spaces. Concretely, we leverage a pre-trained two-character diffusion model to simplify the multi-character interaction space. \textcolor{black}{We group multiple characters by their indices to select corresponding historical motions as conditions for generation.} The diffusion model generates the interactions for each group autoregressively. Additionally, by dividing multiple characters into two-character groups, our multi-character interaction space is agnostic to the number of characters.

The second component, as shown in Fig. \ref{fig:planning} predicts a future transition plan $C^t$ based on the observed interactions $M^{t-1}_{1:N}$. We design the plan to be high-level re-grouping choices for all characters. By re-grouping characters, the newly grouped characters are coordinated to interact with each other in the next motion clip. We leverage a virtual character in a two-character group if the transition plan for this group contains only one existing character.

Specifically, our method follows an autoregressive conditional generative formulation:
\begin{equation}
    M^{t}_{1:N}=\mathcal{F}_\theta(M^{t-1}_{1:N},\epsilon^t_{1:N}, C^t),
\end{equation}
where $\mathcal{F}$ represents our autoregressive generative pipeline, $\epsilon^t_{1:N}$ is the sampled standard Gaussian noise for generation, $M^{t-1}_{1:N}$ is the last observed interaction clip for all characters $1,2\cdots,N$, $M^t_{1:N}$ is the next interaction clip for all characters, $C^t$ is the transition plan for next motion clip $M^t_{1:N}$, and $\theta$ represents all trainable parameters.

\subsection{Coordinatable Multi-Character Interaction Synthesis}\label{sec:interaction_synthesis}

A natural and coordinatable interaction space for multiple characters is vital for coordinated interaction synthesis. It requires the interaction space to have a realistic interaction manifold and to be controllable for coordination. Without available data, we consider simplification by a divide-and-conquer paradigm.

We propose to simplify the interaction space for multiple characters by dividing multiple characters into two-character groups. Multiple character interactions are usually composed of several groups that contain fewer characters \cite{jeong2024multi} and each group can be modeled from an existing model. Theoretically, any division is feasible if a group can be modeled. For our implementation, we divide characters into two-character groups and leverage an existing two-character diffusion model to generate two-character interactions. Through our division, we leverage its ability to synthesize natural interactions for each group. Meanwhile, the division facilitates integrating the transition plan as the conditional signal by re-grouping two characters in an existing group for new interaction groups. Additionally, group division allows our interaction space to be flexible to synthesize varying numbers of characters.

\begin{figure}[htbp]
  \centering
  \includegraphics[width=\linewidth]{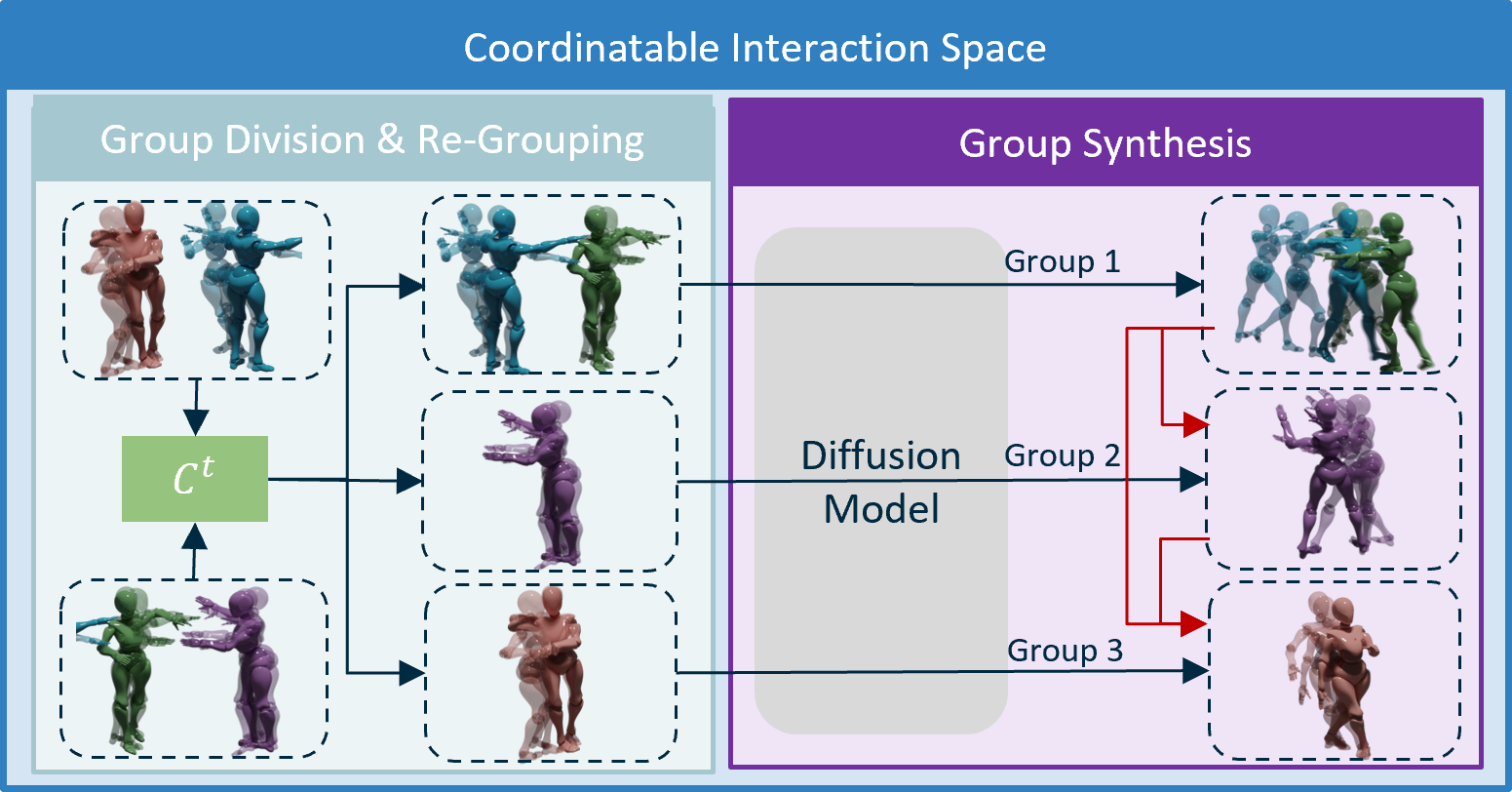}
  \caption{Coordinatable multi-character interaction space by group division. We divide multiple characters into groups and re-group them for potential coordination. The group synthesis generates new motions group by group. The newly generated group is conditioned on the already generated ones, which is indicated by red arrows.}
  \Description{Approximation for multi-character interaction space.}
  \label{fig:approximation}
\end{figure}

We propose to generalize a two-character diffusion model for multiple characters by autoregressively synthesizing groups. We design our generative method to be based on a two-character diffusion model due to its high capacity to model complex data space \cite{chang2023design} and flexible conditional mechanism without being trained on the conditional data \cite{dhariwal2021diffusion}. To autoregressively synthesize the next motion clip for a two-character group, we leverage classifier guidance from the already generated other groups and the observed motion clip for this group. As shown in Fig. \ref{fig:approximation}, the red arrow indicates that we condition on the clips of other groups already generated.

Specifically, when given the observed interaction clip for multiple characters $M^{t-1}_{1:N}$, we generalize a state-of-the-art diffusion model \cite{liang2024intergen} for two-character interactions to generate the next interaction clip for all characters. We autoregressively generate a new social group under the condition of existing social groups. These social groups form the next interaction clip for multiple characters $M^t_{1:N}$. This generation process is formulated as:
\begin{equation}
    M^t_{1:N}=\left[g(M^{t-1}_{i,j},M^\prime),\;for\,(i,j)\,in\,C^t\right],
\end{equation}
where $g$ is the two-character diffusion model, $M^\prime$ represents groups that have already been generated during this autoregressive procedure, and $M^{t-1}_{i,j}$ represent the observed interaction clip for the $i$-th and $j$-th character in a group, $C^t$ is the transition plan to re-grouping observed motions.

We propose to leverage the classifier guidance in the diffusion model to maintain the social distance between the newly generated group $M^t_{i,j}$ and the already generated groups $M^\prime$. To maintain suitable distances, we leverage a distance function $d(\cdot)$ between different groups. \textcolor{black}{The hip distance follows Proxemics Theory \cite{rios2015proxemics}, which states that humans regulate their personal space based on hip position. This promotes the appropriate distances between individuals. Other joints are less representative and would require the pre-trained diffusion model to solve a more complex high-dimensional spatiotemporal constraint.} The distance function measures the average of hip-wise distances:
\begin{equation}
    d(M^t_{i,j},M^\prime)=\frac{1}{|M^\prime|}\sum_{n^\prime}^{|M^\prime|}\min\left(\|p_{i,j}-p_{n^\prime}\|_2^2-\tau, 0\right),
\end{equation}
where $\tau$ is the threshold, and $p_{i,j}$ is the hip position of the character $i$ and $j$. We also constraint the smoothness between the observed clip and the generated clip, which is calculated as:
\begin{equation}
    d(M^t_{i,j},M^{t-1}_{i,j})=\sum\|acc_{i,j}\|_2^2,
\end{equation}
where $acc$ represents the acceleration of all joints. Therefore, the constraint function is:
\begin{equation}
    d=d(M^t_{i,j},M^\prime)+d(M^t_{i,j},M^{t-1}_{i,j})
\end{equation}
Our design for the coordinatable interaction space is summarized in Algorithm \ref{alg:intra}.

\subsection{Transition Planning}

Multi-character interactions contain not only realistic interaction details but also the ability to plan suitable transitions. Their transition is planned on the basis of the currently observed interactions of themselves and others. After a transition is planned, they perform the next actions, which are synthesized by the aforementioned interaction space. Therefore, transition planning contains high-level decisions on interaction groups for multiple characters.

\begin{algorithm}
\caption{Coordinate-able Multi-Character Interaction Space}\label{alg:intra}
\KwData{Re-grouping choice $(i,j)$ from a transition plan $C^t$, motion mask $m$ for motion inpainting, observed motions $M^{t-1}_{1:N}$, other groups $M^\prime$, a pre-trained model $g(\cdot)$, distance function $d(\cdot)$}
\KwResult{a group $M^t_{i,j}$}
$M^{t-1}_{i,j}\gets M^{t-1}_{1:N}[(i,j)]$ \Comment{Re-grouping by transition plan}\;
$u \gets U$         \Comment{$U$ is total number of diffusion timestep}\;
$\epsilon^t\gets \mathcal{N}(0,I)$            \Comment{Sample a random noise}\;
$x^u\gets\epsilon^t$\;
\While{$u \neq 0$}{
    $x^0=g(x^u,u)$    \Comment{Diffusion predicts xstart}\;
    $x^0=m\otimes M^{t-1}_{i,j}+(1-m)\otimes x^0]$   \Comment{Masking for inpainting}\;
    $x^0=x^0+\nabla_x d$   \Comment{Classifier guidance}\;
    $u \gets u - 1$\;
}
$M^t_{i,j}\gets x^0$\;
$M^\prime\gets M^\prime\cup M^t_{i,j}$
\end{algorithm}

To achieve the planning ability, we design a conditional mechanism via re-grouping for the previously introduced multi-character interaction space. While considering all characters for coordination is theoretically feasible, we implement transition planning locally within four characters to pursue a balance between the complexity of grouping choices and the number of potential transition candidates, \textcolor{black}{i.e., the trade-off between transition representativeness and learning complexity. A character evaluating all others as potential transition partners improves representativeness, but increases training complexity. Conversely, considering only the nearest character removes the need for learning, but limits transition diversity.}

The condition signal in our case is the indices of characters to be grouped together. \textcolor{black}{To address the lack of ground truth data, we use re-grouping as high-level semantic control, which avoids the computational burden of determining low-level joint movements in a high-dimensional spatiotemporal space.} Thus, our planning network is agnostic to motion types. Specifically, the planning network takes the motions of considered characters and returns a re-grouping choice as a transition plan, which is formulated as:
\begin{equation}
    C^t=f_{\theta}(M^t_{i,j,i^\prime,j^\prime}).
\end{equation}

\begin{figure}[htbp]
  \centering
  \includegraphics[width=\linewidth]{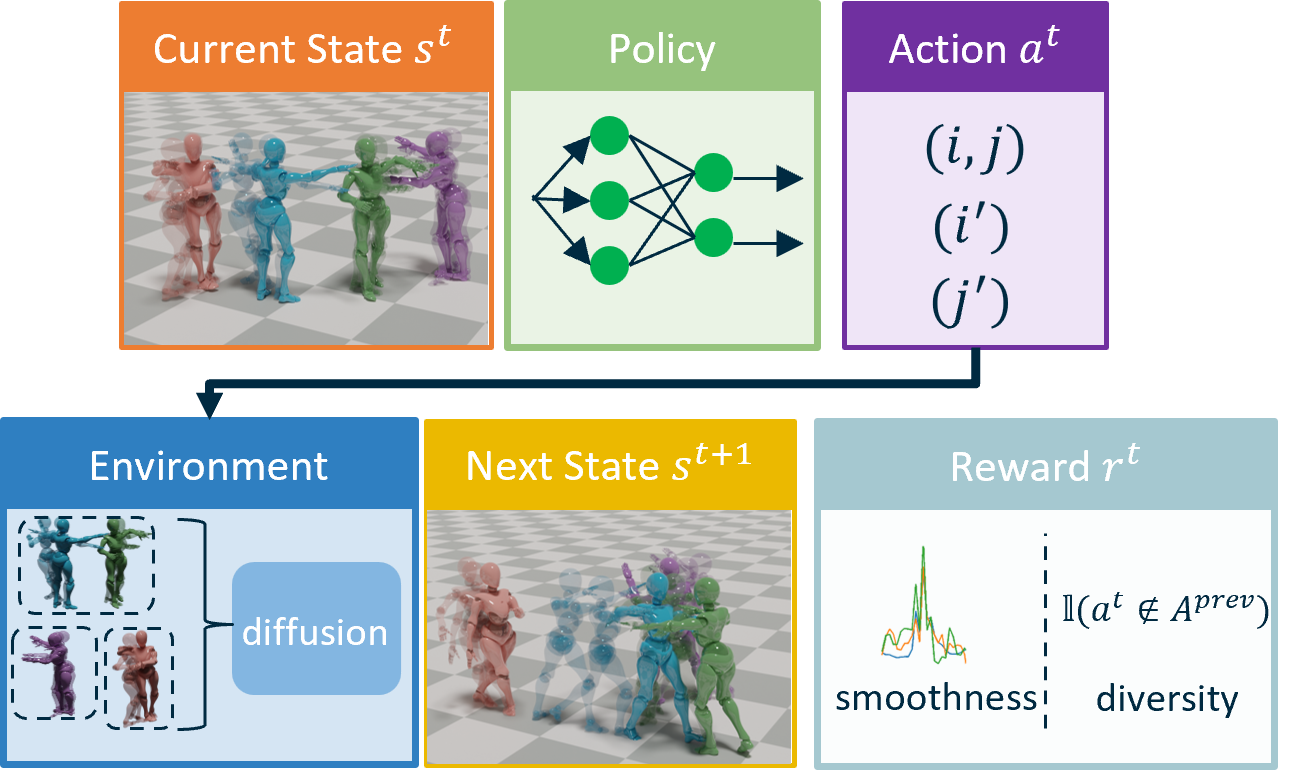}
  \caption{The planning network is learned as a policy network via deep reinforcement learning. The action is a transition plan that contains a high-level grouping choice.}
  \Description{The planning network is learned as a policy network via deep reinforcement learning. The action is a transition plan that contains a high-level grouping choice.}
  \label{fig:planning}
\end{figure}

Due to the lack of an available dataset for our goal, we formulate transition planning as a Markov Decision Process and learn our planning network with a deep reinforcement framework, as depicted in Figure \ref{fig:planning}. The MDP is represented by the tuple $(\mathcal{S},\mathcal{A},\mathcal{R},E)$ where $\mathcal{S}$ is the state space, $\mathcal{A}$ is the action space, $\mathcal{R}$ is a scalar reward function, and $E$ is the environment model.

Specifically, a state $s\in\mathcal{S}$ in our setting consists of clips of characters of interest:
\begin{equation}
    s:=M_{i,j,i^\prime,j^\prime}.
\end{equation}
In our case, an action $a\in\mathcal{A}$ is defined to be a transition plan:
\begin{equation}
    a:=C,
\end{equation}
where $C$ represents the re-grouping choice for the four characters.

We leverage the previously defined interaction space as the environment for reinforcement learning. The transition function $E:(s^t,a^t)\rightarrow s^{t+1}$ describes the probability that we transition to state $s^{t+1}\in\mathcal{S}$ given that we executed action $a^t\in\mathcal{A}$ in state $s^t\in\mathcal{S}$. The interaction synthesis module introduced in Section \ref{sec:interaction_synthesis} works as the environment model in our reinforcement learning setting:
\begin{equation}
    E:=g(\cdot).
\end{equation}
To serve as the environment model, our generative model takes as inputs the current state $M^t$ and generates the next state $M^{t+1}$:
\begin{equation}
    s^{t+1}=g\left(\epsilon^t,s^{t}\right),
\end{equation}
where $\epsilon^t\sim\mathcal{N}(0,I)$. The first state $M^1$ is generated by setting the current state to be empty.

The planning network is formulated as the policy network:
\begin{equation}
    \pi:=f_{\theta}(\cdot).
\end{equation}
It takes four characters' motions as the current state and predicts the next state of these characters:
\begin{equation}
    a^t=\pi(s^t).
\end{equation}

A reward function $R:\mathcal{S}\times\mathcal{A}\times\mathcal{S}\rightarrow R$ and the reward $r:=R(s^t,a^t,s^{t+1})$ evaluate a transition $(s^t,a^t,s^{t+1})$ given the agent task. In our case, the task is to interact with all the other three characters as smoothly as possible. The reward function is designed to advocate for such considerations. The smoothness reward is defined as:
\begin{equation}
    r_{smooth}=e^{-\|acc^{t}-acc^{t+1}}\|^2_2,
\end{equation}
where $acc$ corresponds to the acceleration of motions during the transition. In particular, we consider ten frames to calculate the acceleration. The second one is the diversity to encourage more different transition choices. This diversity is calculated as:
\begin{equation}
    r_{div}=
    \begin{cases}
        1,  & \text{if $a^t$ is novel} \\
        0, & \text{otherwise}
    \end{cases}
\end{equation}
Therefore, we have the following reward function:
\begin{equation}
    r=r_{smooth}+ r_{div}.
\end{equation}

\section{Experiments}

We run a series of experiments to (i) investigate the effectiveness of our method by employing two metrics and (ii) validate the scalability and transferability of our method to three applications.

We evaluate our method with transition smoothness (TS) \cite{barquero2024seamless} and hip distance (HD) for comparison. Following \cite{barquero2024seamless}, transition smoothness calculates the change of acceleration, and we report its maximum change (peak jerk) as a metric. For hip distance, we calculate the average distance between the hip of a character and those of other characters. This metric is designed to indicate whether characters overlapped in the same position. We report the average value over all characters and frames, which is calculated by
\begin{equation}
    HD = \frac{2}{N(N-1)F}\sum_{f=1}^F\sum_{i,j\in N}\|h_i^f-h_j^f\|^2_2,
\end{equation}
where $N$ is the set of all characters, $F$ is the number of frames, $\frac{2}{N(N-1)F}$ is an averaging term, and $h_i^f$ and $h_j^f$ are the hip positions for characters $i$ and $j$ at frame $f$.

\subsection{Quantitative Comparison}

As currently there is no method for implementing our task, we use InterGen \cite{liang2024intergen} as the baseline for our comparison, which is the state-of-the-art method for interaction synthesis. We also implement an adaptation to InterGen for our comparison, and the adapted InterGen is denoted as InterGen$\dagger$. For our adaptation, we implement exactly the same coordinatable interaction space to allow InterGen to generate multiple characters. For transition planning, as InterGen does not have such planning, we randomly sample coordinations and use them as the control signal. To calculate the metrics, we repeat the generation and report the average value. Table \ref{tab:quantitative} shows the results of our comparison. Our method has the best transition smoothness, which indicates that the characters generated by our method transit with fewer artifacts.

\begin{table}[htbp]
\caption{Comparison with interaction synthesis models. $\dagger$ represents our implementation of the coordinatable interaction space in the original method. TS denotes transition smoothness and HD, the hip distance.}
\label{tab:quantitative}
    \centering
    \begin{tabular}{ccc}
    \toprule
        Methods & TS $\downarrow$ & HD \\
    \midrule
        InterGen & 0.073 & 0.567 \\
        InterGen$\dagger$ & 0.117 &  1.578\\
        Ours & \textbf{0.071} & 1.963  \\
    \bottomrule
    \end{tabular}
\end{table}

\begin{figure}[htbp]
  \centering
  \includegraphics[width=0.9\linewidth]{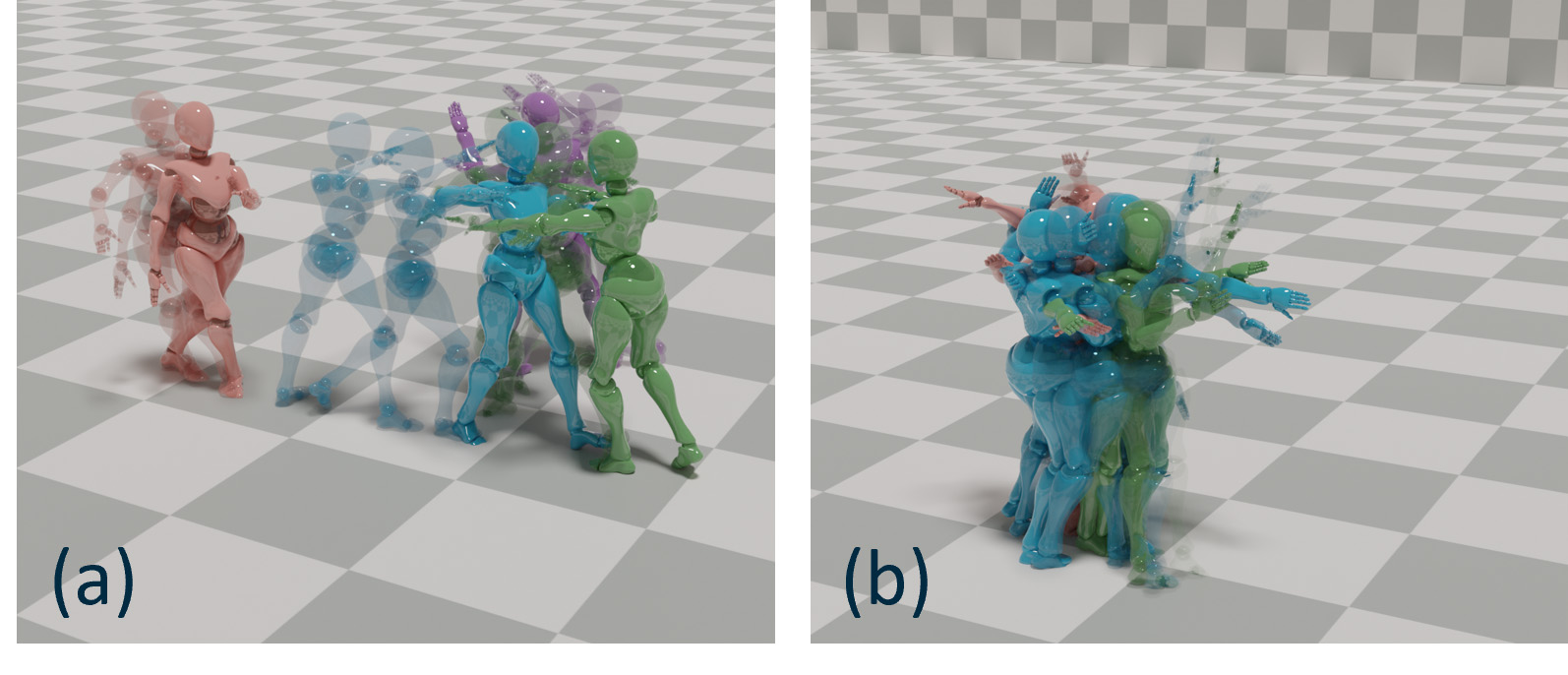}
  \caption{(a) An example result from our method. (b) An example from InterGen where characters heavily overlap.  }
  \Description{(a) An example result from our method. (b) An example from InterGen where characters are overlapped.}
  \label{fig:overlap}
\end{figure}

The InterGen demonstrates close TS performance compared with ours. However, the generated characters are heavily overlapped with each other and the resulting motion is visibly worse. We hypothesize that this better transition smoothness score achieved by InterGen, compared to ours, is because the transition occurs within a much smaller distance between characters and thus the TS value becomes smaller. This is also further corroborated by the hip distance being much higher in InterGen than ours. The InterGen Figure~\ref{fig:overlap} shows that the characters heavily overlap each other.

\begin{figure}[htbp]
  \centering
  \includegraphics[width=\linewidth]{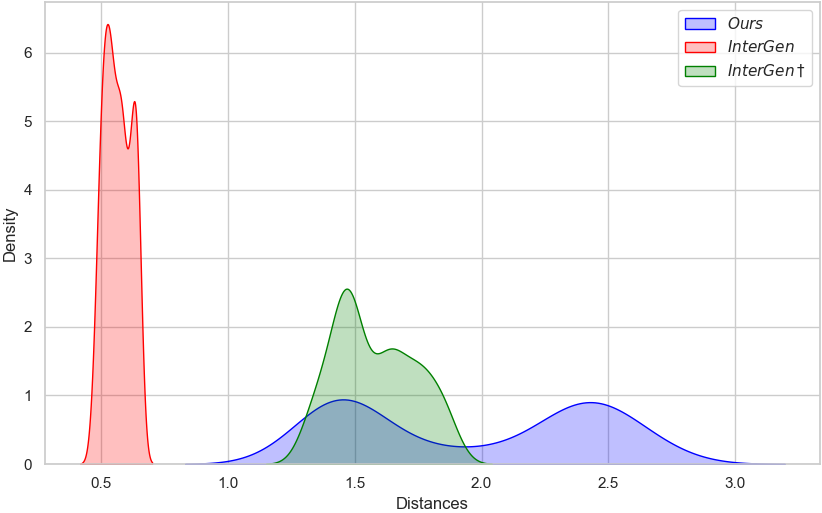}
  \caption{The density of hip distance for the three methods evaluated. The two modes in our hip distance density demonstrate minimal character overlap and clear transitions. InterGen$\dagger$ does not have the ability of transition planning, leading to an averaged distance density with a single mode. InterGen has a similar curve shape with InterGen$\dagger$ as both of them do not have transition planning. Its much smaller mode value indicates that characters heavily overlap.}
  \Description{The density of hip distance for the three methods evaluated. The two modes in our hip distance density demonstrate minimal character overlap and clear transitions. InterGen$\dagger$ does not have the ability of transition planning, leading to an averaged distance density with a single mode. InterGen has a similar curve shape with InterGen$\dagger$ as both of them do not have transition planning. Its much smaller mode value indicates that characters heavily overlap.}
  \label{fig:interaction}
\end{figure}

Figure \ref{fig:interaction} shows the density of the HD values for the three methods. The two modes in our density indicate that the characters are not overlapped and are clearly transited with the learned planning network. InterGen$\dagger$ does not have the ability to plan transitions, which leads to an averaged distance density with only one mode. InterGen has a curve shape similar to that of InterGen$\dagger$ because both do not have transition planning. The mode value of InterGen is much smaller than the other two, which indicates that the characters are heavily overlapped.

As our method generalizes InterGen with coordinatable interaction space and transition planning, Table \ref{tab:quantitative} also shows that the two proposed components are effective. When compared with InterGen$\dagger$, our method performs better in transition planning. When we compare InterGen and InterGen$\dagger$, both of which do not have transition planning, we find that the coordinatable interaction space in InterGen$\dagger$ helps to reduce the overlap in InterGen.

\subsection{Extended Applications}

We demonstrate the scalability of our method and the transferability of our planning network by implementing three applications. Our method is scalable to multiple characters despite only having access to a two-character dataset. We show that synthesizing more characters via our method can be done by either gradually adding new characters or generating a larger scene as a whole. Our planning network is also transferable to other types of motion because our action space is defined as the high-level grouping decision. We show this transferability by taking boxing motions as an example.

\subsubsection{Adding New Character Synthesis}

Our method facilitates the application of adding new characters to achieve a larger number of characters. In this experiment, we start with four characters and gradually add other characters. The newly added characters can also be coordinated to interact with existing characters via the re-grouping choice from our planning network. Table \ref{tab:application} shows the performance of adding new characters.

\subsubsection{Generating Large Scenes}

We generalize our method to synthesize a large number of characters. Although we only have access to a two-character dataset, we generate a lot more characters in one scene. Instead of gradually adding new characters, we choose to generate these characters at once. Table \ref{tab:application} shows the performance in the generation of large scenes.

\subsubsection{Other types of motion}

As we design the planning network to predict the suitable re-grouping choice without considering the actual movements, the planning network is transferable to other motion types. For our experimental implementation, we report the performance of transferring the planning network trained on dancing motions to boxing motions without any adaptation. Table \ref{tab:application} shows the transition smoothness and the hip distance.

When applying our planning network to boxing motions, the transition smoothness does not get degraded and the hip distance increases a bit. This shows that by designing transition plan to be high-level re-grouping operation, the planning network is not related to future movements of characters. This helps our planning network to be transferable. The hip distance increases slightly because boxing motions usually have larger distances than dancing.

\begin{table}[htbp]
\caption{Method performance on extended applications. TS denotes transition smoothness and HD, the hip distance.}
\label{tab:application}
    \centering
    \begin{tabular}{ccc}
    \toprule
        Methods & TS  & HD  \\
    \midrule
        Adding New Characters& 0.026 & 2.450 \\
        Generating Large Scenes & 0.075 & 3.372 \\
        Boxing & 0.057 & 2.155\\
    \bottomrule
    \end{tabular}
\end{table}

\section{Conclusion}

We propose a method to synthesize coordinated interactions for multiple characters without any multi-character dataset. We decompose coordinated interactions into interaction synthesis and coordination planning, and propose a coordinatable interaction space and a planning network, respectively. Considering the lack of data, a pre-trained two-character interaction diffusion model is leveraged in our interaction space by decomposing multi-character interactions into interactions of two-character groups. The planning network provides control signals by re-grouping the observed motions. Our experiments show that the motions from our method are smoother, and character limbs penetrate less with one another compared with other methods. We also demonstrate scalability and transferability with extended applications.

While our method can achieve multi-character coordinated interactions without accessing any multi-character dataset, limitations exist, most of which arise due to the lack of multi-character datasets. The first one is that our method relies on the dividing approximation between multiple characters and two characters. The second one is the controllability of diffusion models when there is no such conditioned dataset. Although classifier guidance and motion inpainting are used in our method, control accuracy can be further improved \cite{karunratanakul2023guided} for better generation quality. Additionally, our method relies on the pre-trained two-character interaction diffusion model to synthesize motions for two-character groups. The synthesis quality of two-character groups also influences the quality of our method. \textcolor{black}{One potential direction for future work is collecting data on close interactions between two or more people. Existing crowd simulation data mainly focus on 2D and recently 3D motions like collision avoidance. With data collected, our method would enhance fine-grained interactions and improve visual realism. For large scenes, an imaginary character is introduced if the total number of characters is not divisible by four and future improvements could include predefined actions such as walking or recycling characters \cite{shum2008interaction} to further enhance transitions. Currently, a 4-character group is formed by a greedy distance-based strategy, and future work could explore other advanced methods such as a first-person receptive field.}

\begin{acks}
This research is supported in part by the EPSRC NortHFutures project (ref: EP/X031012/1).
\end{acks}

\bibliographystyle{ACM-Reference-Format}
\bibliography{main} 


\begin{thebibliography}{39}


\ifx \showCODEN    \undefined \def \showCODEN     #1{\unskip}     \fi
\ifx \showDOI      \undefined \def \showDOI       #1{#1}\fi
\ifx \showISBNx    \undefined \def \showISBNx     #1{\unskip}     \fi
\ifx \showISBNxiii \undefined \def \showISBNxiii  #1{\unskip}     \fi
\ifx \showISSN     \undefined \def \showISSN      #1{\unskip}     \fi
\ifx \showLCCN     \undefined \def \showLCCN      #1{\unskip}     \fi
\ifx \shownote     \undefined \def \shownote      #1{#1}          \fi
\ifx \showarticletitle \undefined \def \showarticletitle #1{#1}   \fi
\ifx \showURL      \undefined \def \showURL       {\relax}        \fi
\providecommand\bibfield[2]{#2}
\providecommand\bibinfo[2]{#2}
\providecommand\natexlab[1]{#1}
\providecommand\showeprint[2][]{arXiv:#2}

\bibitem[Barquero et~al\mbox{.}(2024)]%
        {barquero2024seamless}
\bibfield{author}{\bibinfo{person}{German Barquero}, \bibinfo{person}{Sergio Escalera}, {and} \bibinfo{person}{Cristina Palmero}.} \bibinfo{year}{2024}\natexlab{}.
\newblock \showarticletitle{Seamless Human Motion Composition with Blended Positional Encodings}. In \bibinfo{booktitle}{\emph{Proceedings of the IEEE/CVF Conference on Computer Vision and Pattern Recognition}}.
\newblock


\bibitem[Chang et~al\mbox{.}(2023a)]%
        {chang23unifying}
\bibfield{author}{\bibinfo{person}{Ziyi Chang}, \bibinfo{person}{Edmund J.~C. Findlay}, \bibinfo{person}{Haozheng Zhang}, {and} \bibinfo{person}{Hubert P.~H. Shum}.} \bibinfo{year}{2023}\natexlab{a}.
\newblock \showarticletitle{Unifying Human Motion Synthesis and Style Transfer with Denoising Diffusion Probabilistic Models}. In \bibinfo{booktitle}{\emph{Proceedings of the 2023 International Conference on Computer Graphics Theory and Applications}} (Lisbon, Portugal) \emph{(\bibinfo{series}{GRAPP '23})}. \bibinfo{publisher}{SciTePress}, \bibinfo{pages}{64--74}.
\newblock
\showISBNx{978-989-758-634-7}
\showISSN{2184-4321}
\urldef\tempurl%
\url{https://doi.org/10.5220/0011631000003417}
\showDOI{\tempurl}


\bibitem[Chang et~al\mbox{.}(2023b)]%
        {chang2023design}
\bibfield{author}{\bibinfo{person}{Ziyi Chang}, \bibinfo{person}{George~A Koulieris}, {and} \bibinfo{person}{Hubert~PH Shum}.} \bibinfo{year}{2023}\natexlab{b}.
\newblock \showarticletitle{On the design fundamentals of diffusion models: A survey}.
\newblock \bibinfo{journal}{\emph{arXiv preprint arXiv:2306.04542}} (\bibinfo{year}{2023}).
\newblock


\bibitem[Charalambous et~al\mbox{.}(2023)]%
        {charalambous2023greil}
\bibfield{author}{\bibinfo{person}{Panayiotis Charalambous}, \bibinfo{person}{Julien Pettre}, \bibinfo{person}{Vassilis Vassiliades}, \bibinfo{person}{Yiorgos Chrysanthou}, {and} \bibinfo{person}{Nuria Pelechano}.} \bibinfo{year}{2023}\natexlab{}.
\newblock \showarticletitle{GREIL-Crowds: Crowd Simulation with Deep Reinforcement Learning and Examples}.
\newblock \bibinfo{journal}{\emph{ACM Transactions on Graphics (TOG)}} \bibinfo{volume}{42}, \bibinfo{number}{4} (\bibinfo{year}{2023}), \bibinfo{pages}{1--15}.
\newblock


\bibitem[Dhariwal and Nichol(2021)]%
        {dhariwal2021diffusion}
\bibfield{author}{\bibinfo{person}{Prafulla Dhariwal} {and} \bibinfo{person}{Alexander Nichol}.} \bibinfo{year}{2021}\natexlab{}.
\newblock \showarticletitle{Diffusion models beat gans on image synthesis}.
\newblock \bibinfo{journal}{\emph{Advances in neural information processing systems}}  \bibinfo{volume}{34} (\bibinfo{year}{2021}), \bibinfo{pages}{8780--8794}.
\newblock


\bibitem[Fiore et~al\mbox{.}(2013)]%
        {fiore2013toward}
\bibfield{author}{\bibinfo{person}{Stephen~M Fiore}, \bibinfo{person}{Travis~J Wiltshire}, \bibinfo{person}{Emilio~JC Lobato}, \bibinfo{person}{Florian~G Jentsch}, \bibinfo{person}{Wesley~H Huang}, {and} \bibinfo{person}{Benjamin Axelrod}.} \bibinfo{year}{2013}\natexlab{}.
\newblock \showarticletitle{Toward understanding social cues and signals in human--robot interaction: effects of robot gaze and proxemic behavior}.
\newblock \bibinfo{journal}{\emph{Frontiers in psychology}}  \bibinfo{volume}{4} (\bibinfo{year}{2013}), \bibinfo{pages}{859}.
\newblock


\bibitem[Guo et~al\mbox{.}(2022)]%
        {guo2022multi}
\bibfield{author}{\bibinfo{person}{Wen Guo}, \bibinfo{person}{Xiaoyu Bie}, \bibinfo{person}{Xavier Alameda-Pineda}, {and} \bibinfo{person}{Francesc Moreno-Noguer}.} \bibinfo{year}{2022}\natexlab{}.
\newblock \showarticletitle{Multi-person extreme motion prediction}. In \bibinfo{booktitle}{\emph{Proceedings of the IEEE/CVF Conference on Computer Vision and Pattern Recognition}}. \bibinfo{pages}{13053--13064}.
\newblock


\bibitem[Jeong et~al\mbox{.}(2024)]%
        {jeong2024multi}
\bibfield{author}{\bibinfo{person}{Jaewoo Jeong}, \bibinfo{person}{Daehee Park}, {and} \bibinfo{person}{Kuk-Jin Yoon}.} \bibinfo{year}{2024}\natexlab{}.
\newblock \showarticletitle{Multi-agent long-term 3d human pose forecasting via interaction-aware trajectory conditioning}. In \bibinfo{booktitle}{\emph{Proceedings of the IEEE/CVF Conference on Computer Vision and Pattern Recognition}}. \bibinfo{pages}{1617--1628}.
\newblock


\bibitem[Karunratanakul et~al\mbox{.}(2023)]%
        {karunratanakul2023guided}
\bibfield{author}{\bibinfo{person}{Korrawe Karunratanakul}, \bibinfo{person}{Konpat Preechakul}, \bibinfo{person}{Supasorn Suwajanakorn}, {and} \bibinfo{person}{Siyu Tang}.} \bibinfo{year}{2023}\natexlab{}.
\newblock \showarticletitle{Guided Motion Diffusion for Controllable Human Motion Synthesis}. In \bibinfo{booktitle}{\emph{Proceedings of the IEEE/CVF International Conference on Computer Vision (ICCV)}}. \bibinfo{pages}{2151--2162}.
\newblock


\bibitem[Kim et~al\mbox{.}(2012)]%
        {kim2012tiling}
\bibfield{author}{\bibinfo{person}{Manmyung Kim}, \bibinfo{person}{Youngseok Hwang}, \bibinfo{person}{Kyunglyul Hyun}, {and} \bibinfo{person}{Jehee Lee}.} \bibinfo{year}{2012}\natexlab{}.
\newblock \showarticletitle{Tiling motion patches}. In \bibinfo{booktitle}{\emph{Proceedings of the ACM SIGGRAPH/Eurographics Symposium on Computer Animation}}. \bibinfo{pages}{117--126}.
\newblock


\bibitem[Kwon et~al\mbox{.}(2008)]%
        {kwon2008two}
\bibfield{author}{\bibinfo{person}{Taesoo Kwon}, \bibinfo{person}{Young-Sang Cho}, \bibinfo{person}{Sang~I Park}, {and} \bibinfo{person}{Sung~Yong Shin}.} \bibinfo{year}{2008}\natexlab{}.
\newblock \showarticletitle{Two-character motion analysis and synthesis}.
\newblock \bibinfo{journal}{\emph{IEEE transactions on visualization and computer graphics}} \bibinfo{volume}{14}, \bibinfo{number}{3} (\bibinfo{year}{2008}), \bibinfo{pages}{707--720}.
\newblock


\bibitem[Li et~al\mbox{.}(2023)]%
        {li2023object}
\bibfield{author}{\bibinfo{person}{Jiaman Li}, \bibinfo{person}{Jiajun Wu}, {and} \bibinfo{person}{C~Karen Liu}.} \bibinfo{year}{2023}\natexlab{}.
\newblock \showarticletitle{Object motion guided human motion synthesis}.
\newblock \bibinfo{journal}{\emph{ACM Transactions on Graphics (TOG)}} \bibinfo{volume}{42}, \bibinfo{number}{6} (\bibinfo{year}{2023}), \bibinfo{pages}{1--11}.
\newblock


\bibitem[Liang et~al\mbox{.}(2024)]%
        {liang2024intergen}
\bibfield{author}{\bibinfo{person}{Han Liang}, \bibinfo{person}{Wenqian Zhang}, \bibinfo{person}{Wenxuan Li}, \bibinfo{person}{Jingyi Yu}, {and} \bibinfo{person}{Lan Xu}.} \bibinfo{year}{2024}\natexlab{}.
\newblock \showarticletitle{Intergen: Diffusion-based multi-human motion generation under complex interactions}.
\newblock \bibinfo{journal}{\emph{International Journal of Computer Vision}} (\bibinfo{year}{2024}), \bibinfo{pages}{1--21}.
\newblock


\bibitem[Liu et~al\mbox{.}(2006)]%
        {liu2006composition}
\bibfield{author}{\bibinfo{person}{C~Karen Liu}, \bibinfo{person}{Aaron Hertzmann}, {and} \bibinfo{person}{Zoran Popovi{\'c}}.} \bibinfo{year}{2006}\natexlab{}.
\newblock \showarticletitle{Composition of complex optimal multi-character motions}. In \bibinfo{booktitle}{\emph{Proceedings of the 2006 ACM SIGGRAPH/Eurographics symposium on Computer animation}}. \bibinfo{pages}{215--222}.
\newblock


\bibitem[Mao et~al\mbox{.}(2023)]%
        {mao2023leapfrog}
\bibfield{author}{\bibinfo{person}{Weibo Mao}, \bibinfo{person}{Chenxin Xu}, \bibinfo{person}{Qi Zhu}, \bibinfo{person}{Siheng Chen}, {and} \bibinfo{person}{Yanfeng Wang}.} \bibinfo{year}{2023}\natexlab{}.
\newblock \showarticletitle{Leapfrog diffusion model for stochastic trajectory prediction}. In \bibinfo{booktitle}{\emph{Proceedings of the IEEE/CVF Conference on Computer Vision and Pattern Recognition}}. \bibinfo{pages}{5517--5526}.
\newblock


\bibitem[Panayiotou et~al\mbox{.}(2022)]%
        {panayiotou2022ccp}
\bibfield{author}{\bibinfo{person}{Andreas Panayiotou}, \bibinfo{person}{Theodoros Kyriakou}, \bibinfo{person}{Marilena Lemonari}, \bibinfo{person}{Yiorgos Chrysanthou}, {and} \bibinfo{person}{Panayiotis Charalambous}.} \bibinfo{year}{2022}\natexlab{}.
\newblock \showarticletitle{Ccp: Configurable crowd profiles}. In \bibinfo{booktitle}{\emph{ACM SIGGRAPH 2022 conference proceedings}}. \bibinfo{pages}{1--10}.
\newblock


\bibitem[Pelechano et~al\mbox{.}(2016)]%
        {pelechano2016simulating}
\bibfield{author}{\bibinfo{person}{Nuria Pelechano}, \bibinfo{person}{Jan~M Allbeck}, \bibinfo{person}{Mubbasir Kapadia}, {and} \bibinfo{person}{Norman~I Badler}.} \bibinfo{year}{2016}\natexlab{}.
\newblock \bibinfo{booktitle}{\emph{Simulating heterogeneous crowds with interactive behaviors}}.
\newblock \bibinfo{publisher}{CRC Press}.
\newblock


\bibitem[Petrovich et~al\mbox{.}(2022)]%
        {petrovich2022temos}
\bibfield{author}{\bibinfo{person}{Mathis Petrovich}, \bibinfo{person}{Michael~J Black}, {and} \bibinfo{person}{G{\"u}l Varol}.} \bibinfo{year}{2022}\natexlab{}.
\newblock \showarticletitle{TEMOS: Generating diverse human motions from textual descriptions}. In \bibinfo{booktitle}{\emph{European Conference on Computer Vision}}. Springer, \bibinfo{pages}{480--497}.
\newblock


\bibitem[Rios-Martinez et~al\mbox{.}(2015)]%
        {rios2015proxemics}
\bibfield{author}{\bibinfo{person}{Jorge Rios-Martinez}, \bibinfo{person}{Anne Spalanzani}, {and} \bibinfo{person}{Christian Laugier}.} \bibinfo{year}{2015}\natexlab{}.
\newblock \showarticletitle{From proxemics theory to socially-aware navigation: A survey}.
\newblock \bibinfo{journal}{\emph{International Journal of Social Robotics}}  \bibinfo{volume}{7} (\bibinfo{year}{2015}), \bibinfo{pages}{137--153}.
\newblock


\bibitem[Shum et~al\mbox{.}(2008b)]%
        {shum2008interaction}
\bibfield{author}{\bibinfo{person}{Hubert~PH Shum}, \bibinfo{person}{Taku Komura}, \bibinfo{person}{Masashi Shiraishi}, {and} \bibinfo{person}{Shuntaro Yamazaki}.} \bibinfo{year}{2008}\natexlab{b}.
\newblock \showarticletitle{Interaction patches for multi-character animation}.
\newblock \bibinfo{journal}{\emph{ACM transactions on graphics (TOG)}} \bibinfo{volume}{27}, \bibinfo{number}{5} (\bibinfo{year}{2008}), \bibinfo{pages}{1--8}.
\newblock


\bibitem[Shum et~al\mbox{.}(2007)]%
        {shum2007simulating}
\bibfield{author}{\bibinfo{person}{Hubert~PH Shum}, \bibinfo{person}{Taku Komura}, {and} \bibinfo{person}{Shuntaro Yamazaki}.} \bibinfo{year}{2007}\natexlab{}.
\newblock \showarticletitle{Simulating competitive interactions using singly captured motions}. In \bibinfo{booktitle}{\emph{Proceedings of the 2007 ACM symposium on Virtual reality software and technology}}. \bibinfo{pages}{65--72}.
\newblock


\bibitem[Shum et~al\mbox{.}(2008a)]%
        {shum2008simulating}
\bibfield{author}{\bibinfo{person}{Hubert~PH Shum}, \bibinfo{person}{Taku Komura}, {and} \bibinfo{person}{Shuntaro Yamazaki}.} \bibinfo{year}{2008}\natexlab{a}.
\newblock \showarticletitle{Simulating interactions of avatars in high dimensional state space}. In \bibinfo{booktitle}{\emph{Proceedings of the 2008 Symposium on interactive 3D Graphics and Games}}. \bibinfo{pages}{131--138}.
\newblock


\bibitem[Shum et~al\mbox{.}(2010)]%
        {shum2010simulating}
\bibfield{author}{\bibinfo{person}{Hubert~PH Shum}, \bibinfo{person}{Taku Komura}, {and} \bibinfo{person}{Shuntaro Yamazaki}.} \bibinfo{year}{2010}\natexlab{}.
\newblock \showarticletitle{Simulating multiple character interactions with collaborative and adversarial goals}.
\newblock \bibinfo{journal}{\emph{IEEE Transactions on Visualization and Computer Graphics}} \bibinfo{volume}{18}, \bibinfo{number}{5} (\bibinfo{year}{2010}), \bibinfo{pages}{741--752}.
\newblock


\bibitem[Starke et~al\mbox{.}(2022)]%
        {starke2022deepphase}
\bibfield{author}{\bibinfo{person}{Sebastian Starke}, \bibinfo{person}{Ian Mason}, {and} \bibinfo{person}{Taku Komura}.} \bibinfo{year}{2022}\natexlab{}.
\newblock \showarticletitle{Deepphase: Periodic autoencoders for learning motion phase manifolds}.
\newblock \bibinfo{journal}{\emph{ACM Transactions on Graphics (TOG)}} \bibinfo{volume}{41}, \bibinfo{number}{4} (\bibinfo{year}{2022}), \bibinfo{pages}{1--13}.
\newblock


\bibitem[Tevet et~al\mbox{.}(2022a)]%
        {tevet2022motionclip}
\bibfield{author}{\bibinfo{person}{Guy Tevet}, \bibinfo{person}{Brian Gordon}, \bibinfo{person}{Amir Hertz}, \bibinfo{person}{Amit~H Bermano}, {and} \bibinfo{person}{Daniel Cohen-Or}.} \bibinfo{year}{2022}\natexlab{a}.
\newblock \showarticletitle{Motionclip: Exposing human motion generation to clip space}. In \bibinfo{booktitle}{\emph{European Conference on Computer Vision}}. Springer, \bibinfo{pages}{358--374}.
\newblock


\bibitem[Tevet et~al\mbox{.}(2022b)]%
        {tevet2022human}
\bibfield{author}{\bibinfo{person}{Guy Tevet}, \bibinfo{person}{Sigal Raab}, \bibinfo{person}{Brian Gordon}, \bibinfo{person}{Yonatan Shafir}, \bibinfo{person}{Daniel Cohen-Or}, {and} \bibinfo{person}{Amit~H Bermano}.} \bibinfo{year}{2022}\natexlab{b}.
\newblock \showarticletitle{Human motion diffusion model}.
\newblock \bibinfo{journal}{\emph{arXiv preprint arXiv:2209.14916}} (\bibinfo{year}{2022}).
\newblock


\bibitem[Won et~al\mbox{.}(2021)]%
        {won2021control}
\bibfield{author}{\bibinfo{person}{Jungdam Won}, \bibinfo{person}{Deepak Gopinath}, {and} \bibinfo{person}{Jessica Hodgins}.} \bibinfo{year}{2021}\natexlab{}.
\newblock \showarticletitle{Control strategies for physically simulated characters performing two-player competitive sports}.
\newblock \bibinfo{journal}{\emph{ACM Transactions on Graphics (TOG)}} \bibinfo{volume}{40}, \bibinfo{number}{4} (\bibinfo{year}{2021}), \bibinfo{pages}{1--11}.
\newblock


\bibitem[Won et~al\mbox{.}(2014)]%
        {won2014generating}
\bibfield{author}{\bibinfo{person}{Jungdam Won}, \bibinfo{person}{Kyungho Lee}, \bibinfo{person}{Carol O'Sullivan}, \bibinfo{person}{Jessica~K Hodgins}, {and} \bibinfo{person}{Jehee Lee}.} \bibinfo{year}{2014}\natexlab{}.
\newblock \showarticletitle{Generating and ranking diverse multi-character interactions}.
\newblock \bibinfo{journal}{\emph{ACM Transactions on Graphics (TOG)}} \bibinfo{volume}{33}, \bibinfo{number}{6} (\bibinfo{year}{2014}), \bibinfo{pages}{1--12}.
\newblock


\bibitem[Xiang et~al\mbox{.}(2024)]%
        {xiang2024socialcvae}
\bibfield{author}{\bibinfo{person}{Wei Xiang}, \bibinfo{person}{YIN Haoteng}, \bibinfo{person}{He Wang}, {and} \bibinfo{person}{Xiaogang Jin}.} \bibinfo{year}{2024}\natexlab{}.
\newblock \showarticletitle{SocialCVAE: Predicting Pedestrian Trajectory via Interaction Conditioned Latents}. In \bibinfo{booktitle}{\emph{Proceedings of the AAAI Conference on Artificial Intelligence}}, Vol.~\bibinfo{volume}{38}. \bibinfo{pages}{6216--6224}.
\newblock


\bibitem[Xie et~al\mbox{.}(2024)]%
        {xie2024omnicontrol}
\bibfield{author}{\bibinfo{person}{Yiming Xie}, \bibinfo{person}{Varun Jampani}, \bibinfo{person}{Lei Zhong}, \bibinfo{person}{Deqing Sun}, {and} \bibinfo{person}{Huaizu Jiang}.} \bibinfo{year}{2024}\natexlab{}.
\newblock \showarticletitle{OmniControl: Control Any Joint at Any Time for Human Motion Generation}. In \bibinfo{booktitle}{\emph{The Twelfth International Conference on Learning Representations}}.
\newblock


\bibitem[Xu et~al\mbox{.}(2023b)]%
        {xu2023actformer}
\bibfield{author}{\bibinfo{person}{Liang Xu}, \bibinfo{person}{Ziyang Song}, \bibinfo{person}{Dongliang Wang}, \bibinfo{person}{Jing Su}, \bibinfo{person}{Zhicheng Fang}, \bibinfo{person}{Chenjing Ding}, \bibinfo{person}{Weihao Gan}, \bibinfo{person}{Yichao Yan}, \bibinfo{person}{Xin Jin}, \bibinfo{person}{Xiaokang Yang}, {et~al\mbox{.}}} \bibinfo{year}{2023}\natexlab{b}.
\newblock \showarticletitle{ActFormer: A GAN-based transformer towards general action-conditioned 3D human motion generation}. In \bibinfo{booktitle}{\emph{Proceedings of the IEEE/CVF International Conference on Computer Vision}}. \bibinfo{pages}{2228--2238}.
\newblock


\bibitem[Xu et~al\mbox{.}(2024)]%
        {xu2024regennet}
\bibfield{author}{\bibinfo{person}{Liang Xu}, \bibinfo{person}{Yizhou Zhou}, \bibinfo{person}{Yichao Yan}, \bibinfo{person}{Xin Jin}, \bibinfo{person}{Wenhan Zhu}, \bibinfo{person}{Fengyun Rao}, \bibinfo{person}{Xiaokang Yang}, {and} \bibinfo{person}{Wenjun Zeng}.} \bibinfo{year}{2024}\natexlab{}.
\newblock \showarticletitle{Regennet: Towards human action-reaction synthesis}. In \bibinfo{booktitle}{\emph{Proceedings of the IEEE/CVF Conference on Computer Vision and Pattern Recognition}}. \bibinfo{pages}{1759--1769}.
\newblock


\bibitem[Xu et~al\mbox{.}(2023a)]%
        {xu2023joint}
\bibfield{author}{\bibinfo{person}{Qingyao Xu}, \bibinfo{person}{Weibo Mao}, \bibinfo{person}{Jingze Gong}, \bibinfo{person}{Chenxin Xu}, \bibinfo{person}{Siheng Chen}, \bibinfo{person}{Weidi Xie}, \bibinfo{person}{Ya Zhang}, {and} \bibinfo{person}{Yanfeng Wang}.} \bibinfo{year}{2023}\natexlab{a}.
\newblock \showarticletitle{Joint-Relation Transformer for Multi-Person Motion Prediction}. In \bibinfo{booktitle}{\emph{Proceedings of the IEEE/CVF International Conference on Computer Vision}}. \bibinfo{pages}{9816--9826}.
\newblock


\bibitem[Yersin et~al\mbox{.}(2009)]%
        {yersin2009crowd}
\bibfield{author}{\bibinfo{person}{Barbara Yersin}, \bibinfo{person}{Jonathan Ma{\"\i}m}, \bibinfo{person}{Julien Pettr{\'e}}, {and} \bibinfo{person}{Daniel Thalmann}.} \bibinfo{year}{2009}\natexlab{}.
\newblock \showarticletitle{Crowd patches: populating large-scale virtual environments for real-time applications}. In \bibinfo{booktitle}{\emph{Proceedings of the 2009 symposium on Interactive 3D graphics and games}}. \bibinfo{pages}{207--214}.
\newblock


\bibitem[Yuan et~al\mbox{.}(2023)]%
        {yuan2023physdiff}
\bibfield{author}{\bibinfo{person}{Ye Yuan}, \bibinfo{person}{Jiaming Song}, \bibinfo{person}{Umar Iqbal}, \bibinfo{person}{Arash Vahdat}, {and} \bibinfo{person}{Jan Kautz}.} \bibinfo{year}{2023}\natexlab{}.
\newblock \showarticletitle{Physdiff: Physics-guided human motion diffusion model}. In \bibinfo{booktitle}{\emph{Proceedings of the IEEE/CVF International Conference on Computer Vision}}. \bibinfo{pages}{16010--16021}.
\newblock


\bibitem[Yue et~al\mbox{.}(2024)]%
        {yue2024human}
\bibfield{author}{\bibinfo{person}{Jiangbei Yue}, \bibinfo{person}{Baiyi Li}, \bibinfo{person}{Julien Pettr{\'e}}, \bibinfo{person}{Armin Seyfried}, {and} \bibinfo{person}{He Wang}.} \bibinfo{year}{2024}\natexlab{}.
\newblock \showarticletitle{Human Motion Prediction Under Unexpected Perturbation}. In \bibinfo{booktitle}{\emph{Proceedings of the IEEE/CVF Conference on Computer Vision and Pattern Recognition}}. \bibinfo{pages}{1501--1511}.
\newblock


\bibitem[Zhang et~al\mbox{.}(2023)]%
        {zhang2023simulation}
\bibfield{author}{\bibinfo{person}{Yunbo Zhang}, \bibinfo{person}{Deepak Gopinath}, \bibinfo{person}{Yuting Ye}, \bibinfo{person}{Jessica Hodgins}, \bibinfo{person}{Greg Turk}, {and} \bibinfo{person}{Jungdam Won}.} \bibinfo{year}{2023}\natexlab{}.
\newblock \showarticletitle{Simulation and retargeting of complex multi-character interactions}. In \bibinfo{booktitle}{\emph{ACM SIGGRAPH 2023 Conference Proceedings}}. \bibinfo{pages}{1--11}.
\newblock


\bibitem[Zhong et~al\mbox{.}(2022)]%
        {zhong2022data}
\bibfield{author}{\bibinfo{person}{Jinghui Zhong}, \bibinfo{person}{Dongrui Li}, \bibinfo{person}{Zhixing Huang}, \bibinfo{person}{Chengyu Lu}, {and} \bibinfo{person}{Wentong Cai}.} \bibinfo{year}{2022}\natexlab{}.
\newblock \showarticletitle{Data-driven crowd modeling techniques: A survey}.
\newblock \bibinfo{journal}{\emph{ACM Transactions on Modeling and Computer Simulation (TOMACS)}} \bibinfo{volume}{32}, \bibinfo{number}{1} (\bibinfo{year}{2022}), \bibinfo{pages}{1--33}.
\newblock


\bibitem[Zhou et~al\mbox{.}(2023)]%
        {zhou2023multi}
\bibfield{author}{\bibinfo{person}{Kanglei Zhou}, \bibinfo{person}{Hubert~PH Shum}, \bibinfo{person}{Frederick~WB Li}, {and} \bibinfo{person}{Xiaohui Liang}.} \bibinfo{year}{2023}\natexlab{}.
\newblock \showarticletitle{Multi-task spatial-temporal graph auto-encoder for hand motion denoising}.
\newblock \bibinfo{journal}{\emph{IEEE Transactions on Visualization and Computer Graphics}} \bibinfo{volume}{30}, \bibinfo{number}{10} (\bibinfo{year}{2023}), \bibinfo{pages}{6754--6769}.
\newblock


\end{thebibliography}

\begin{figure*}
\centering
    \includegraphics[width=\linewidth]{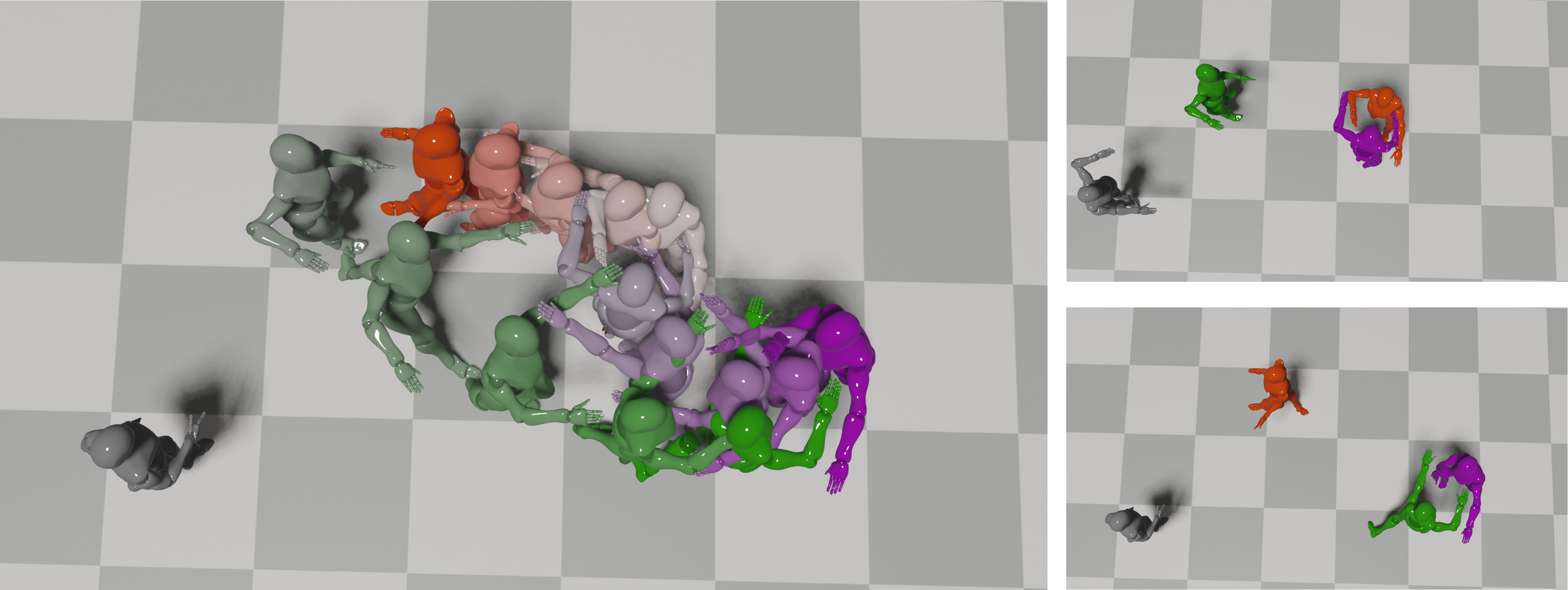}
    \caption{Snapshot of key frames during a transition period for four characters. The more saturated the color, the more recent the frame. We highlight the three interacting characters in red, purple, and green, while the others are grey as this transition only involves the three colored characters. (Left) We show character movements during the transition using key frames. In this transition, the purple character previously interacted with the red one and now is coordinated to interact with the green one. (Upper right) The starting key frame of this transition where the red and purple characters are interacting with each other. (Lower right) The ending key frame of this transition where the purple and green characters are interacting with each other.}
    \Description{Snapshot of key frames during a transition period for four characters. The more saturated the color, the more recent the frame. We highlight the three interacting characters in red, purple, and green, while the others are grey as this transition only involves the three colored characters. (Left) We show character movements during the transition using key frames. In this transition, the purple character previously interacted with the red one and now is coordinated to interact with the green character. (Upper right) The starting key frame of this transition where the red and purple characters are interacting with each other. (Lower right) The ending key frame of this transition where the purple and green characters are interacting with each other.}
    \label{fig:four-figure}
\end{figure*}

\begin{figure*}
\centering
    \includegraphics[width=\linewidth]{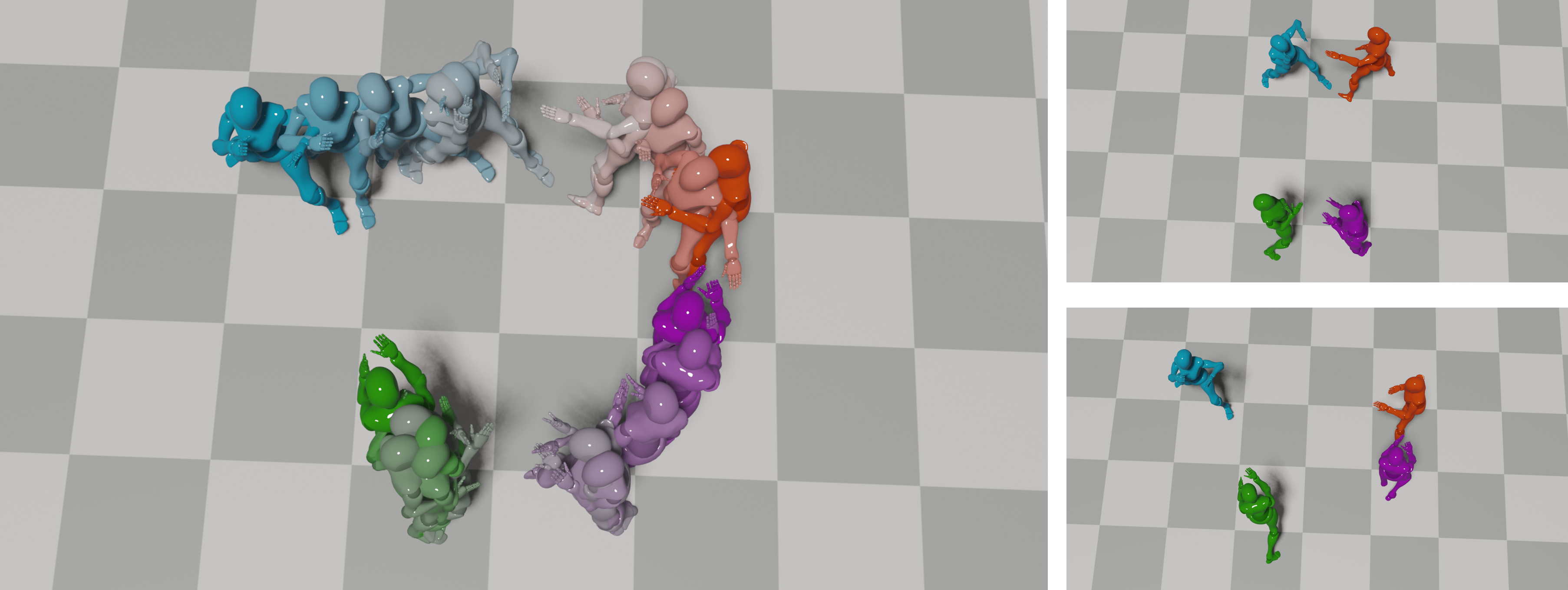}
    \caption{Snapshot of key frames during a transition period in the experiment of transferring to a different motion type, e.g., boxing. The more saturated the color, the more recent the frame. We highlight the four characters in red, purple, blue, and green. (Left) We show character movements during the transition using key frames. In this transition, the purple and green characters previously interacted with each other, as did the red and blue ones. Now the red and purple characters are coordinated to interact, and as do the blue and green ones. (Upper right) The starting key frame of this transition where the red and blue characters are interacting with each other, and the purple and green ones are also interacting with each other. (Lower right) The ending key frame of this transition where the purple and red characters are interacting with each other, and the blue and green are interacting with each other.}
    \Description{Snapshot of key frames during a transition period in the experiment of transferring to a different motion type, e.g., boxing. The more saturated the color, the more recent the frame. We highlight the four characters in red, purple, blue, and green. (Left) We show character movements during the transition using key frames. In this transition, the purple and green characters previously interacted with each other, as did the red and blue ones. Now the red and purple characters are coordinated to interact, and as do the blue and green ones. (Upper right) The starting key frame of this transition where the red and blue characters are interacting with each other, and the purple and green ones are also interacting with each other. (Lower right) The ending key frame of this transition where the purple and red characters are interacting with each other, and the blue and green are interacting with each other.}
    \label{fig:transfer-figure}
\end{figure*}

\begin{figure*}
\centering
    \includegraphics[width=\linewidth]{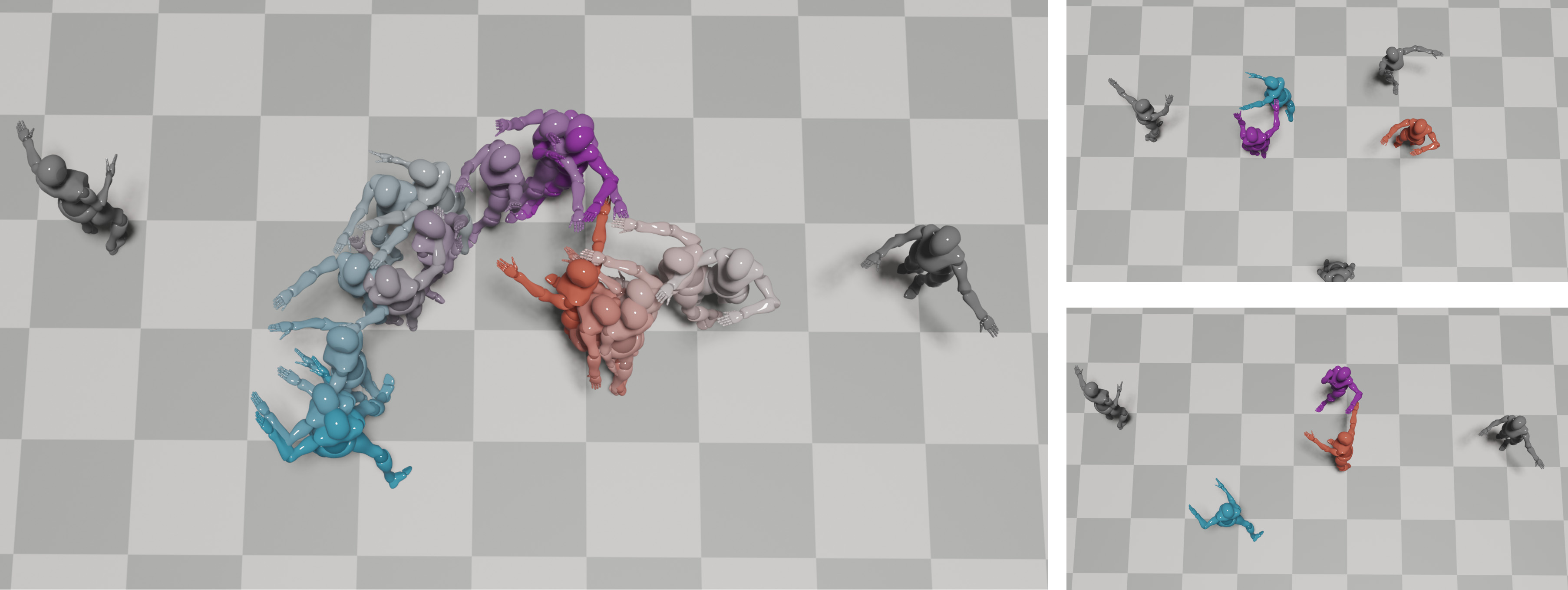}
    \caption{Snapshot of key frames during a transition period in the experiment of adding new characters. The more saturated the color, the more recent the frame. We highlight the three characters in red, purple, and blue while the others are grey because this transition only involves the three colored characters. The purple and blue characters are added more recently than the red character. (Left) We show their movements during the transition using key frames. In this transition, the purple character previously interacted with the blue one and now is coordinated to interact with the red character. (Upper right) The starting key frame of this transition where the blue and purple characters are interacting with each other. (Lower right) The ending key frame of this transition where the purple and red characters are interacting with each other.}
    \Description{Snapshot of key frames during a transition period in the experiment of adding new characters. The more saturated the color, the more recent the frame. We highlight the three characters in red, purple, and blue while the others are grey because this transition only involves the three colored characters. The purple and blue characters are added more recently than the red character. (Left) We show their movements during the transition using key frames. In this transition, the purple character previously interacted with the blue one and now is coordinated to interact with the red character. (Upper right) The starting key frame of this transition where the blue and purple characters are interacting with each other. (Lower right) The ending key frame of this transition where the purple and red characters are interacting with each other.}
    \label{fig:adding-figure}
\end{figure*}

\begin{figure*}
\centering
    \includegraphics[width=\linewidth]{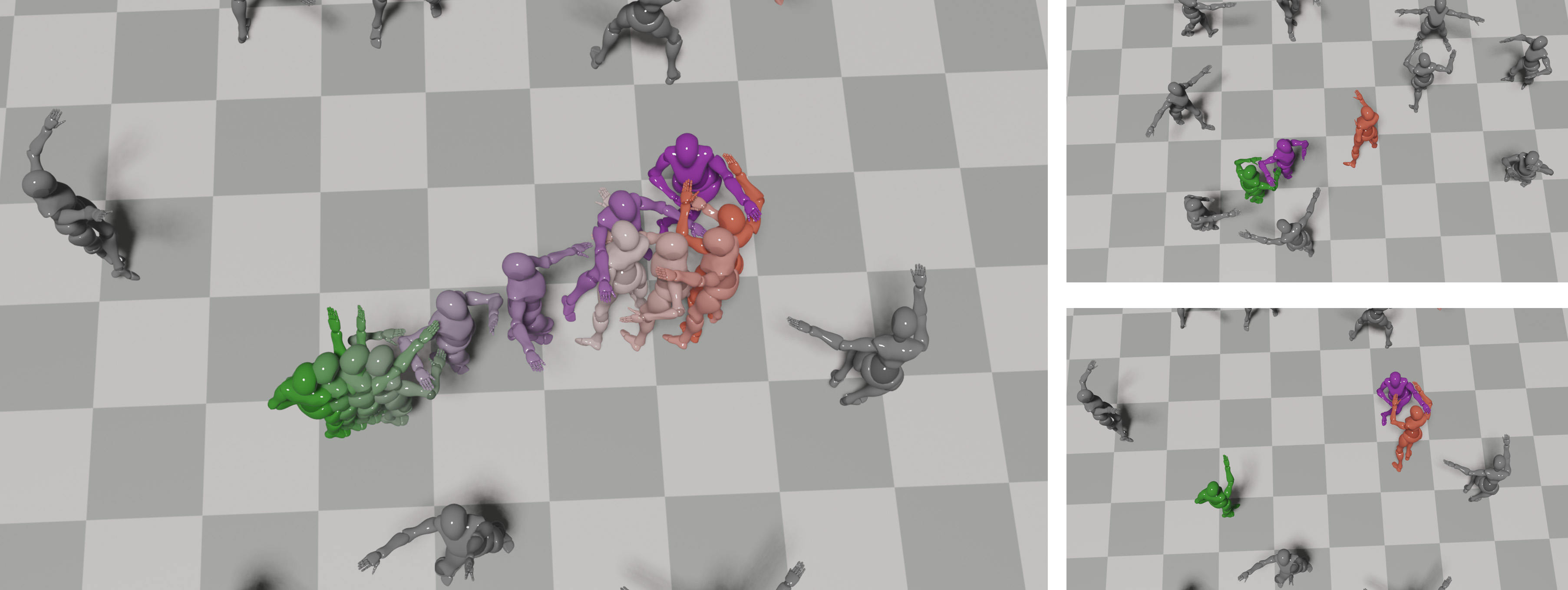}
    \caption{Snapshot of key frames during a transition period in the experiment involving a large number of characters. The more saturated the color, the more recent the frame. We highlight the three characters in red, purple, and green, while the others are grey because this transition only involves the three colored characters. (Left) We show their movements during the transition using key frames. In this transition, the purple character previously interacted with the green one and now is coordinated to interact with the red character. (Upper right) The starting key frame of this transition where the green and purple characters are interacting with each other. (Lower right) The ending key frame of this transition where the purple and red characters are interacting with each other.}
    \Description{Snapshot of key frames during a transition period in the experiment involving a large number of characters. The more saturated the color, the more recent the frame. We highlight the three characters in red, purple, and green, while the others are grey because this transition only involves the three colored characters. (Left) We show their movements during the transition using key frames. In this transition, the purple character previously interacted with the green one and now is coordinated to interact with the red character. (Upper right) The starting key frame of this transition where the green and purple characters are interacting with each other. (Lower right) The ending key frame of this transition where the purple and red characters are interacting with each other.}
    \label{fig:large-figure}
\end{figure*}

\clearpage


\section*{Supplementary material}

# Large-Scale Multi-Character Interaction Synthesis Supplementary Material

Ziyi Chang
ziyi.chang@durham.ac.uk
Durham University
Durham, United Kingdom

He Wang
he_wang@ucl.ac.uk
University College London
London, United Kingdom

George Alex Koulieris
georgios.a.koulieris@durham.ac.uk
Durham University
Durham, United Kingdom

Hubert P. H. Shum*
hubert.shum@durham.ac.uk
Durham University
Durham, United Kingdom

## 1 TWO-CHARACTER GROUP DIVISION CHOICE

We clarify that our two-character interaction setup is a modeling choice driven by two factors: (1) Unlike prior work that loosely couples single-character motions, we focus on close and continuous interactions (CCI) with frequent, prolonged contact (e.g., dancing), which rarely involve more than two people. (2) Data availability is a major constraint. Capturing two-person CCI is already rare, and, to our knowledge, no datasets exist for multi-person CCI.

Our model is not inherently limited to two-person interactions. To demonstrate this, we adapt our method using Multi-Track Timeline Control [Petrovich et al. 2024], allowing a character to be in multiple groups (e.g., A-B, A-C) with minimal modifications: diffusion model taking two interaction groups as input and further constraining distances among three individuals. Results are shown in Table 1:

Table 1: **Allowing three-character division choice.**

|  | TS | HD |
| --- | --- | --- |
| Two-character division | 0.071 | 1.963 |
| Allowing three-character division | 0.079 | 1.888 |

This confirms our method's high extensibility with minimal modifications, enabling diverse scenarios and demonstrating potential for broader multi-character interaction synthesis.

## 2 EVALUATION METRICS

### 2.1 User Study

We provide a small user study where participants were asked to rank the quality of results generated by our method against InterGen and InterGen†. 94.12% prefers ours to the other two, 5.88% prefers InterGen†, and InterGen is unfavorable due to heavy character overlap.

### 2.2 Quantitative Metircs

Quantitative metrics remain an open challenge. Following common practice, we measure variance as the diversity metric for dancing, with and without planning. Table 2 shows the diversity results.

*Corresponding author.

Table 2: **Diversity as a quantitative metric.**

|  | Coordinatable Space | Planning | Diversity |
| --- | --- | --- | --- |
| InterGen (dancing) | NA | NA | 4.938 |
| Ours | ✓ | × | 5.001 |
| Ours | ✓ | ✓ | 5.010 |

We attribute the slightly increased diversity to our coordinatable space and planning for synthesizing transitions that are absent in the original two-character data.

Essentially, without directly comparable ground truth, only self-contained metrics can be calculated or measured via quality assessment [Zhou et al. 2023]. Our TS metric has already integrated speed information by considering acceleration changes, which are crucial for physical plausibility and perceptual quality. Biomechanically, smooth acceleration results from the gradual modulation of neural signals regulating muscle force, aligning with Newton's second law.

## 3 HYPER-PARAMETER SETTINGS

Hyper-parameters related to the reinforcement learning part have been shown in Table 3.

Table 3: **Hyperparameters**

| Hyperparameter | Value |
| --- | --- |
| batch size | 16 |
| gamma | 0.98 |
| epsilon start | 0.08 |
| epsilon end | 0.01 |
| epsilon decay | 200 |
| lr | 0.0005 |

DQN strategy is used to train the transition planning network. We used the official diffusion model checkpoints and trained on an NVIDIA RTX 2080 Ti.

## 4 DISCUSSION ON BIASES, RISKS, AND POTENTIAL MITIGATION

Potential biases in motion synthesis can stem from unrepresentative datasets. Some datasets may fail to capture the inherently rich variability of human motions and interactions performed by people from different cultures, genders, and ages. The generated motions and interactions usually concentrate on dominant patterns while under-representing minorities. This limitation not only risks reinforcing stereotypical portrayals and cultural insensitivity but also raises privacy concerns, as models trained on data with skewed distributions may expose sensitive personal information from motion patterns such as genders and ages. Moreover, the under-representation of minority groups can exacerbate the issue, resulting in a loop where certain gestures or movements are insufficiently recognized or reproduced, thereby perpetuating social biases and limiting the generalizability of synthetic models.

To mitigate these challenges, several solutions can be potentially applied. For example, data augmentation for interactions [Li et al. 2024] expands and diversifies training data to better represent the full spectrum of motions and interactions. Another promising solution is diffusion minority sampling [Um et al. 2024]. Sampling strategies have been designed to promote minority patterns contained in the training data to receive adequate emphasis during the generation process. These techniques aim to enhance the fairness of motion synthesis systems, ultimately fostering more inclusive and ethically responsible applications in areas ranging from entertainment and virtual reality to rehabilitation and robotics.

## 5 ABLATION STUDIES ON METHOD SCALING

We performed an ablation study by removing the proposed components to showcase the scaling ability. Table 4 shows the results.

Table 4: **An ablation study on method scaling.**

|      | Coordinatable Space | Planning | TS    | HD    |
|------|---------------------|----------|-------|-------|
| a    | ×                   | ×        | 0.071 | 0.564 |
| b    | ✓                   | ×        | 0.202 | 3.422 |
| Full | ✓                   | ✓        | 0.075 | 3.372 |

## 6 ABLATION STUDIES ON DISTANCE THRESHOLD

We performed an ablation study by changing the distance threshold to demonstrate its effect. The results are shown in Table 5.

## 7 VISUALIZATION ON THE EFFECTS OF TRANSITION SMOOTHNESS METRIC

We provide an example as an illustration of the effects of the transition smoothness metric. Generally speaking, higher values indicate larger discontinuities like abrupt movements, which is indicated by the orange color in Figure 1.

Table 5: **an ablation study on the distance threshold.**

| Distance threshold | TS    | HD    |
|--------------------|-------|-------|
| 0.5                | 0.059 | 1.723 |
| 1.0                | 0.057 | 1.841 |
| 1.5                | 0.067 | 1.901 |
| 2.0                | 0.071 | 1.963 |
| 2.5                | 0.085 | 2.003 |
| 3.0                | 0.092 | 2.355 |

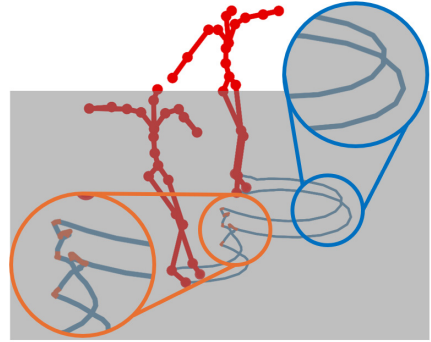

Figure 1: **An illustrative figure for the effects of transition smoothness metric.** Discontinuities in trajectories are highlighted in orange color.

## REFERENCES

Baiyi Li, Edmond SL Ho, Hubert PH Shum, and He Wang. 2024. Two-person interaction augmentation with skeleton priors. In *Proceedings of the IEEE/CVF Conference on Computer Vision and Pattern Recognition*. 1900–1910.

Mathis Petrovich, Or Litany, Umar Iqbal, Michael J Black, Gul Varol, Xue Bin Peng, and Davis Rempe. 2024. Multi-track timeline control for text-driven 3D human motion generation. In *Proceedings of the IEEE/CVF Conference on Computer Vision and Pattern Recognition*. 1911–1921.

Soobin Um, Suhyeon Lee, and Jong Chul Ye. 2024. Don't Play Favorites: Minority Guidance for Diffusion Models. In *The Twelfth International Conference on Learning Representations*. https://openreview.net/forum?id=3NmO9lY4Jn

Kanglei Zhou, Yue Ma, Hubert PH Shum, and Xiaohui Liang. 2023. Hierarchical graph convolutional networks for action quality assessment. *IEEE Transactions on Circuits and Systems for Video Technology* 33, 12 (2023), 7749–7763.



\end{document}